\documentclass[a4paper,11pt]{article}
\usepackage{graphicx}
\usepackage{graphics}
\title{\bf Target mass corrections to matrix elements in  
nucleon spin structure functions}

\author{Y. B. Dong \\
Institute of High Energy Physics, 
CAS, Beijing 100049, P. R. China\\
and\\
Theoretical Physics Center for Science Facilities (TPCSF), CAS, P. R. China}
\begin{document}
%\date{}
\maketitle
\begin{abstract}
Target mass corrections to the twist-4 terms $\tilde f_2^{p,n,d}$ as 
well as to the leading-twist $\tilde a_2$ are discussed. 
 
\end{abstract}

\par\noindent\par\vspace{0.2cm}
We know that different approaches [1-7] have been employed to study 
higher-twist effect to the nucleon structure functions. There were also 
several phenomenological analyses of the nucleon structure functions to 
study quark-hadron duality and to extract the higher-twist contributions 
(like the ones of the twist-3 and twist-4 terms) from experimental 
measurements [8-11]. Those analyses are going to be more and more accurate 
since the more and more precise measurements of the nucleon spin structure 
functions $g_1$ and $g_2$ are becoming available [11-12]. The high precision 
data have been employed to study the validity of the quark-hadron duality 
for the nucleon structure function $F_2$ [13] and even for spin asymmetry 
$A_1$ by HERMES [14] recently. Several experiments to test the higher-twist 
effect on the nucleon spin structure functions are being carried out in the 
Jefferson Laboratory [9,15].

It has been pointed out, in the literature, that the target mass corrections 
(TMCs) should be considered in the studies of the nucleon structure functions 
[16] in a moderate $Q^2$ region, and of the Bloom-Gilman quark-hadron duality 
[17-18]. Therefore, only after the important target mass corrections are 
removed from the experimental data, one can reasonably extract the 
higher-twist effect [18]. There were several papers about the target mass 
corrections to $F_{1,2}(x,Q^2)$ and $g_{1,2}(x,Q^2)$ in the past [19]. 
Recently, the target mass corrections to the nucleon structure functions for 
the polarized deep-inelastic scattering have been systematically studied 
[20-21]. In our previous work [22], TMCs to the twist-3 matrix element in 
the nucleon structure functions are addressed. In this report, TMCs to the 
twist-4 terms $\tilde f_2^{p,n,d}$ as well as to the leading-twist 
$\tilde a_2$ will be discussed. 
 
Consider the Cornwall-Norton (CN) moments 
$g_{1,2}^{(n)}(Q^2)=\int_0^1x^{n-1}g_{1,2}(x,Q^2)dx$, we know that the first 
CN moment of $g_1$ can be generally expanded in inverse powers of $Q^2$ in 
operator production expansion (OPE) [1-2] as
\begin{eqnarray}
g_1^{(1)}=\int_0^1dx g_1(x,Q^2)=\sum_{\tau=2,even}^{\infty}
\frac{\mu_{\tau}(Q^2)}{Q^{\tau-2}}
\end{eqnarray}
with the coefficients $\mu_{\tau}$ relating to the nucleon matrix elements 
of operators of twist $\leq \tau$. In Eq. (1), the leading-twist (twist-2) 
component $\mu_2$ is determined by the matrix elements of the axial vector 
operator $\bar{\psi}\gamma_{\mu}\gamma_5\psi$, summed over various quark 
flavors. The coefficient of $1/Q^2$ term, 
$\mu_4=\frac19M^2(\tilde a_2+4\tilde d_2+4\tilde f_2)$, contains the 
contributions from the twist-2 $\tilde a_2$, twist-3 $\tilde d_2$, and 
twist-4 $\tilde f_2$, respectively. Usually, $\tilde d_2$ is extracted from 
the third moments of the measured $g_1(x,Q^2)$ and $g_2(x,Q^2)$ by using 
$\tilde d_2(Q^2)=\int_0^1 x^2\Big (2g_1(x,Q^2)+3g_2(x,Q^2)\Big )dx$. 
However, it is pointed out that this  method for  $\tilde d_2$ ignores the 
target mass corrections to the third moments of $g_{1,2}$, and the target 
mass corrections play a sizeable role to $\tilde d_2$ [22] in a moderate 
$Q^2$ region.

To further estimate TMCs to the twist-4 of the nucleon spin structure 
functions, one may assume that the contributions from higher-twist term 
with $\tau>6$ can be ignored [23] or assume this term to be a constant 
(neglecting any possible $Q^2$-dependence) [8]. Based on the first 
assumption, we have
\begin{eqnarray}
\frac49y^2\tilde f_2+\frac12\tilde a_0=
g_1^{(1)}-\frac{1}{9}y^2(\tilde a_2+4\tilde d_2).
\end{eqnarray}
When no TMCs are considered, $\tilde a_2$ and $\tilde d_2$ can be simply 
expressed by the CN moments of the nucleon spin structure functions, and we 
get  
\begin{eqnarray}
\frac49y^2\tilde f_2^{(0)}+\frac12\tilde a_0=
g_1^{(1)}-\frac29y^2(5g_1^{(3)}+6g_2^{(3)}). 
\end{eqnarray}

When TMCs are considered, we have to employ the Nachtmann moments 
\begin{eqnarray}
M^{(n)}_1(Q^2)&=&\int^1_0dx\frac{\xi^{n+1}}{x^2}
\bigg \{\Big [\frac{x}{\xi}-\frac{n^2}{(n+2)^2}y^2x\xi
\Big ]g_1(x,Q^2)-y^2x^2\frac{4n}{n+2}g_2(x,Q^2)\bigg \}, 
\nonumber \\
M^{(n)}_2(Q^2)&=&\int^1_0dx\frac{\xi^{n+1}}{x^2}
\bigg \{\frac{x}{\xi}g_1(x,Q^2)
+\Big [\frac{n}{n-1}\frac{x^2}{\xi^2}-
\frac{n}{n+1}y^2x^2\Big ]g_2(x,Q^2)\bigg \},
\end{eqnarray}
where the Nachtmann variable $\xi=\frac{2x}{1+r}$ (with 
$r=\sqrt{1+4y^2x^2}$), $y^2=M^2/Q^2$, and $x$ is the Bjorken variable. 
The two Nachtmann moments are simultaneously constructed by 
the two spin structure functions $g_{1,2}$. If $g_{1,2}(x,Q^2)$ are 
replaced by the ones with TMCs (see Refs. [20-22]), one can easily expand 
the two Nachtmann moments with respect to $y^2$. The results are 
$M_1^{(n)}=\frac12\tilde a_{n-1}+ {\cal O}\Big (y^8\Big )$, and 
$M_2^{(n)}=\frac12\tilde d_{n-1}+{\cal O}\Big (y^8\Big )$.
The two expressions explicitly tell that, different from the CN moments,  
one can get the contributions of a pure twist-2 with spin-n and a pure 
twist-3 with spin-(n-1) operators from the Nachtmann moments. 
The advantage of the Nachtmann moments means that they contain only dynamical 
higher-twist, which are the ones related to the correlations among the 
partons. As a result, they are constructed to protect the 
moments of the nucleon spin structure functions from the target mass 
corrections. Consequently, to extract the higher-twist effect, say twist-3 or 
twist-4 contribution, one is required to consider the Nachtmann moments 
instead of the CN moments. 

We use the Nachtmann moments to express $\tilde a_n$ and $\tilde d_n$ 
and obtain
\begin{eqnarray}
&&\frac49y^2\tilde f_2+\frac12\tilde a_0
=g_1^{(1)}\nonumber \\
&-& \frac29y^2\int_0^1\frac{\xi^4}{x^2}dx
\Big [\Big (\frac{5x}{\xi}-\frac{9}{25}y^2x\xi\Big )g_1(x,Q^2)
+\Big (6\frac{x^2}{\xi^2}-\frac{27}{5}y^2x^2\Big )g_2(x,Q^2)\Big ]
\end{eqnarray}
Thus, TMCs to the twist-4 contribution, due to the two different 
moments, is $\Delta f_2=\tilde f_2-\tilde f_2^0$. Here, we employ the 
parametrization forms of the spin structure functions of the proton, neutron 
and deuteron [11-12] to estimate $\Delta f_2$. 
Note that the well-known Wandzura and Wilczek (WW) relation [24] 
$g_2(x,Q^2)=g_2^{WW}(x,Q^2)=-g_1(x,Q^2)+\int_x^1\frac{g_1(y,Q^2)}{y}dy$
is valid if only the leading-twist is considered, and TMCs to the 
twist-2 contribution do not break the WW relation. However, if the 
higher-twist operators, like twist-3 and twist-4, are considered, the WW 
relation $g_2(x,Q^2)=g_2^{WW}(x,Q^2)$ no longer preserves. Thus, one may 
write $g_2(x,Q^2)=g_2^{WW}(x,Q^2)+\bar{g}_2(x,Q^2)$ [8,9], where 
$\bar{g}_2$ represents the violation of the WW relation. The non-vanishing 
value of $\bar{g}_2$ just results from the higher-twist effect. 

One can calculate $\Delta f_2$ with the parametrizations of $g_{1,2}$. 
The results are plotted in Fig. 1. We see that the typical values 
of the differences are in order of $10^{-3}\sim 10^{-4}$. There are several 
theoretical estimated values for the twist-4 term $\tilde f_2$ in the 
literature (see table 1), like the ones of the bag model [4], of the QCD 
sum rule [5,6], of the empirical analyses of the experimental measurements 
[8, 23], and of the instanton model [25].  Comparing the estimated  
differences in Fig. 1 to those estimated values displayed in table 1, 
we conclude that TMCs to the twist-4 term $\tilde f_2$ 
are negligible (less than 2\%). We also find that $\Delta f_2$ of the 
proton and deuteron are always larger than that of the neutron.

In addition, we check TMCs to the leading twist term (with spin-3) 
$\tilde a_2$. If no TMCs are considered, $\tilde a_2^{(0)}=2 g_1^{(3)}$. 
When TMCs are taken into account, we get, from the Nachtmann moments, 
\begin{eqnarray}
\tilde a_2=\int_0^12\frac{\xi^4}{x^2}dx\Bigg 
\{\Big [\frac{x}{\xi}-\frac{9}{25}y^2x\xi\Big ]g_1(x,Q^2)
-\frac{12}{5}y^2x^2g_2(x,Q^2)\Bigg \}.
\end{eqnarray}  
Fig. 2 displays the $Q^2$-dependence of the ratio 
$R=\tilde a_{2}/\tilde a_2^{(0)}$ for the proton, neutron 
and deuteron targets. The sizable effect of TMCs is clearly seen, since the  
ratios all diverge from unity obviously. When $Q^2\sim 5~GeV^2$, the effect 
of TMCs is still about 10\% for the proton and deuteron targets. In addition, 
the effect on the proton and deuteron targets is much 
larger than that on the neutron. Here the $Q^2$-dependences of the three 
ratios are similar to those of the twist-3 terms [22]. The sizeable effect 
tells that TMCs should be taken into account. Therefore, to estimate the 
matrix element of $\tilde a_2$, the Nachtmann moments are required to be 
employed. 

%{\hskip 1in}
\begin{figure}[t]
\centering
\includegraphics[width=10cm,height=7cm]{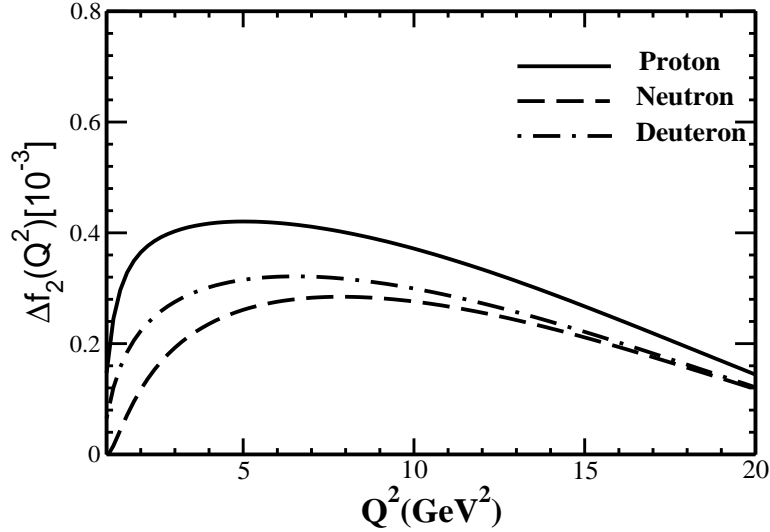}
\caption{\footnotesize Difference $\Delta f_2$
The solid, dashed and dotted-dashed curves are the results of the proton, 
neutron and deuteron, respectively. }
\end{figure}

%{\hskip 1in}
\begin{figure}[t]
\centering
\includegraphics[width=10cm,height=7cm]{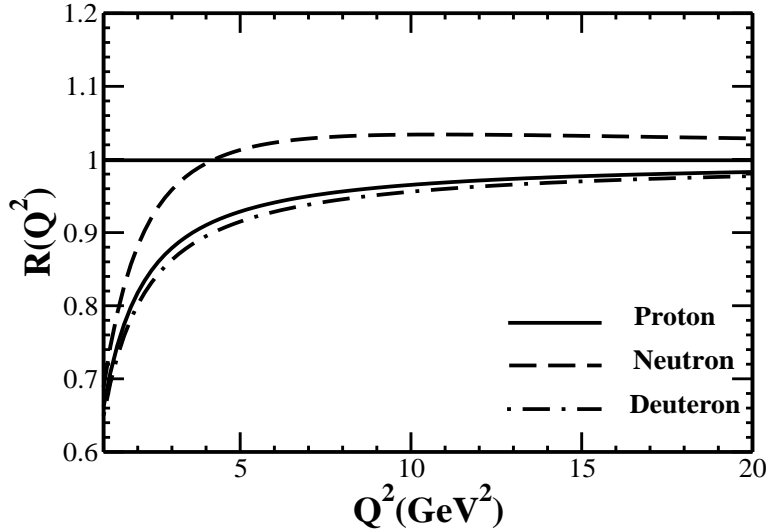}
\caption{\footnotesize Ratio $R_{a_3}$. The solid, dashed and dotted-dashed 
curves are the results of the proton, neutron and deuteron, respectively. }
\end{figure} 

\begin{center}
Table 1, The estimated values for $\tilde f_2$ in different approaches in 
the literature.\\
\vspace{0.5cm}
\begin{tabular}{|c|c|c||c|c|c|}
\hline
References &$\tilde f_2^p$ &$\tilde f_2^n$ 
&References &$\tilde f_2^p$ &$\tilde f_2^n$ \\ \hline
Ref. [4] &$0.050\pm 0.034$ &$-0.018\pm 0.017$
&Ref. [5] &$-0.028$ &$0$ \\ \hline 
Ref. [6] &$0.037\pm 0.006$ &$0.013\pm 0.006$ 
&Ref. [8] &----- &$0.034\pm 0.043$\\ \hline
Ref. [23] &$-0.10\pm 0.05$ &$-0.07\pm 0.08$
&Ref. [25] &$-0.046$ &$0.038$\\ \hline
\end{tabular}
\end{center}

\par\noindent\par\vspace{0.2cm}

In summary, we have explicitly shown the target mass corrections to 
the twist-4 $\tilde f_2$ term and to the leading-twist one (spin-3) 
$\tilde a_2$.  It is reiterated that in order to precisely and consistently 
extract the contributions of the leading-twist $\tilde a_2$, of the twist-3 
$\tilde d_2$ and of the twist-4 $\tilde f_2$ with a definite spin and with a 
moderate $Q^2$ value, one is required to employ the Nachtmann moments 
$M_{1,2}$ instead of the CN moments. Our results show that TMCs play an 
evidently role to  $\tilde a_2$ when $Q^2$ is small. The above conclusion 
does not change if different parameterizations of the structure functions 
are employed. We also show that TMCs to the twist-4 term is much 
smaller than those to the twist-3 term and to the leading-twist term. 

Finally, the expressions of the differences $\Delta f_2$ and $\Delta a_2$
between the CN and Nachtmann moments are 
\begin{eqnarray}
\Delta f_2&=&\tilde f_2-\tilde f_2^{(0)}=\frac{y^2}{10}\Bigg \{ \Big
[\frac{384}{5}g_1^{(5)}-234y^2g_1^{(7)}+736y^4g_1^{(9)}\Big ]
\nonumber \\
&&+\Big [87g_2^{(5)}
-258y^2g_2^{(7)}+798y^4g_2^{(9)}\Big ]\Bigg \}+{\cal O}(y^8),\nonumber \\
\Delta a_2&=&\tilde a_2-\tilde a_2^0=2M^{(3)}_1-2g_1^{(3)}
=y^2\Bigg \{ \Big [-\frac{168}{25}g_1^{(5)}
+\frac{108}{5}y^2g_1^{(7)}-\frac{352}{5}y^4g_1^{(9)}\Big ]\nonumber \\
&&+\Big [-\frac{24}{5}g_2^{(5)}
+\frac{96}{5}y^2g_2^{(7)}-\frac{336}{5}y^4g_2^{(9)}\Big ]\Bigg \}
+{\cal O}(y^8).
\end{eqnarray}
One sees that the two expressions mainly depend on the higher-moment 
of the nucleon spin structure functions, and therefore, on the spin structure 
function in the large-x region. In the most of the empirical analyses of the 
Ellis-Jaffe sum rule (the first moment of $g_1$), the contribution from the 
spin structure function in the large-x region is assumed to be trivial, since 
it behaves like $(1-x)^3$. When the higher-moment of the spin structure 
function is considered, the effect of the spin structure functions in the 
large-x region becomes important. Consequently, the measurement of the 
nucleon spin structure functions in the large-x region with a high precision 
is required. 

\section*{Acknowledgments}
\par\noindent\par\noindent\par 

This work is supported by the National Sciences Foundations of China
under grant No. 10775148, and  by the CAS Knowledge Innovation 
Project No. KJCX3-SYW-N2.

\end{document}